%% file: main.tex
\documentclass[submission,copyright,creativecommons]{eptcs}
\providecommand{\event}{DICE 2017} 
\usepackage{breakurl}             
\usepackage{underscore}           

\usepackage{amsmath}

\usepackage[utf8]{inputenc}
\usepackage[english]{babel}
\usepackage{graphicx}
\usepackage[]{algorithm2e}
\usepackage{tikz}
\usepackage{dot2texi}
\usepackage{listings}
\usepackage{makecell}
\usepackage{fancybox}
\usepackage{tikz}
\usepackage{hyperref}
\usepackage{xcolor}
\usepackage{tabu}
\usepackage{tabto}
\usepackage{colortbl}
\usetikzlibrary{matrix,shapes,arrows,shadows}
\usepackage{amsmath,amssymb,bm,times}

\definecolor{dkgreen}{rgb}{0,0.6,0}
\definecolor{gray}{rgb}{0.5,0.5,0.5}
\definecolor{mauve}{rgb}{0.58,0,0.82}

\newcommand{\comm}[1]{\mathtt{#1}}
\newcommand{\Out}[1]{\mathrm{Out}(#1)}
\newcommand{\In}[1]{\mathrm{In}(#1)}

\newcommand{\dfg}[1]{\mathrm{M}(\comm{#1})}
\newcommand{\limit}{\mathrm{limit}}
\newcommand{\Var}{\mathrm{Var}}

\newcommand{\Code}[1]{\texttt{#1}}

\usepackage{blindtext, graphicx}
\usepackage{listings}
\usepackage{llvm/lang}  
\usepackage{nasm/style} 
\usepackage{amsmath}
\usepackage{tikz}
\usepackage{sidecap}
\usepackage{caption}
\usepackage{subcaption}
\usepackage{tkz-berge}
\usepackage{xcolor}
\usepackage{colortbl}
\usepackage{ragged2e}
\usepackage{booktabs}
\usepackage{stmaryrd}
\usepackage{gnuplottex}
\usetikzlibrary{fit,matrix,shapes,arrows,shadows}
\usepackage{multirow}

\newtheorem{mydef}{Definition}

\newtheorem{mylemma}{Lemma}

\definecolor{myblue}{RGB}{80,80,160}
\definecolor{mygreen}{RGB}{80,160,80}
\definecolor{mygray}{RGB}{80,80,80}

\title{Loop Quasi-Invariant Chunk Motion\\\Large by peeling with
statement composition}
\author{Jean-Yves Moyen
            \institute{Department of Computer Science\\
            University of Copenhagen (DIKU)}
           \email{moyen@lipn.univ-paris13.fr}
           \and
           Thomas Rubiano
           \institute{Universit\'e Paris 13 - LIPN}
            \institute{Department of Computer Science\\
            University of Copenhagen (DIKU)}
           \email{rubiano@lipn.univ-paris13.fr}
           \and
            Thomas Seiller
            \institute{Department of Computer Science\\
            University of Copenhagen (DIKU)}
           \email{seiller@di.ku.dk}
}
\def\titlerunning{Loop Quasi-Invariant Chunk Motion}
\def\authorrunning{J.Y. Moyen, T. Rubiano, T. Seiller}
\begin{document}
\maketitle



\input{intro.tex}
\input{part1.tex}
\input{part2.tex}
\input{conclusion.tex}



\paragraph{Acknowledgments} The authors wish to thank L. Kristiansen
for communicating a manuscript \cite{LarsDraft} that initiated the
present work.  Jean-Yves Moyen is supported by the European
Commision’s Marie Skłodowska-Curie Individual Fellowship
(H2020-MSCA-IF-2014) 655222 - Walgo; Thomas Rubiano is supported by
the ANR project ``Elica'' ANR-14-CE25-0005; Thomas Seiller is
supported by the European Commision’s Marie Skłodowska-Curie
Individual Fellowship (H2020-MSCA-IF-2014) 659920 - ReACT.



\bibliographystyle{eptcs}


\bibliography{main-short}




\end{document}

%% file: intro.tex
\begin{abstract}
  Several techniques for analysis and transformations are used in
  compilers. Among them, the peeling of loops for hoisting
  quasi-invariants can be used to optimize generated code, or simply
  ease developers’ lives. In this paper, we introduce a new concept of
  dependency analysis borrowed from the field of Implicit
  Computational Complexity (ICC), allowing to work with composed
  statements called ``Chunks'' to detect more quasi-invariants. Based
  on an optimization idea given on a \texttt{WHILE} language, we
  provide a transformation method - reusing ICC concepts and
  techniques \cite{LarsDraft,Kuck:1981:DGC:567532.567555} - to
  compilers. This new analysis computes an invariance degree for each
  statement or chunks of statements by building a new kind of
  dependency graph, finds the “maximum” or “worst” dependency graph
  for loops, and recognizes if an entire block is Quasi-Invariant or
  not. This block could be an inner loop, and in that case the
  computational complexity of the overall program can be decreased.
  We already implemented a proof of concept on a toy C
  parser\footnote{\url{https://github.com/ThomasRuby/LQICM\_On\_C\_Toy\_Parser}}
  analysing and transforming the AST representation.

  In this paper, we introduce the theory around this concept and
  present a prototype analysis pass implemented on LLVM. In a very
  near future, we will implement the corresponding transformation and
  provide benchmarks comparisons.
\end{abstract}



\section{Introduction}

A command inside a loop is an \emph{invariant} if its execution has no
effect after the first iteration of the loop.  Typically, an
assignment \texttt{x:=0} in a loop is invariant (provided \texttt{x}
is not modified elsewhere). Loop invariants can safely be moved out of
loops (\emph{hoisted}) in order to make the program faster.

A command inside a loop is \emph{quasi-invariant} if its execution has
no effect after a finite number of iterations of the loop. Typically,
if a loop contains the sequence \texttt{x:=y, y:=0}, then
\texttt{y:=0} is invariant. However, \texttt{x:=y} is \textbf{not}
invariant. The first time the loop is executed, \texttt{x} will be
assigned the old value of \texttt{y}, and only from the second time
onward will \texttt{x} be assigned the value \texttt{0}. Hence, this
command is \emph{quasi-invariant}. It can still be hoisted out of the
loop, but to do so requires to \emph{peel} the loop first, that is
execute its body once (by copying it before the loop).

The number of times a loop must be executed before a quasi-invariant
can be hoisted is called here the \emph{degree} of the invariant.

An obvious way to detect quasi-invariants is to first detect
invariants (that is, quasi-invariants of degree 1) and hoist them; and
iterate the process to find quasi-invariant of degree 2, and so
on. This is, however, not very efficient since it may require a large
number of iterations to find some invariance degrees.

We provide here an analysis able to directly detect the invariance
degree of any statements in the loop. Moreover, our analysis is able
to assign an invariance degree non only to individual statements but
also to groups of statements (called \emph{chuncks}). That way it is
possible, for example, to detect that a whole inner loop is invariant
and hoist it, thus decreasing the asymptotic complexity of the
program.

\smallskip

Loop optimization techniques based on quasi-invariance are well-known
in the compilers community. The transformation idea is to peel loops a
finite number of time and hoist invariants until there are no more
quasi-invariants. As far as we know, this technique is called
``peeling'' and it was introduced by Song \emph{et al.}
\cite{LQICM2000}.

The present paper offers a new point of view on this work. From an
optimization on a \texttt{WHILE} language by Lars Kristiansen
\cite{LarsDraft}, we provide a redefinition of peeling and another
transformation method based on techniques developed in the field of
Implicit Computational Complexity.

Implicit Computational Complexity (ICC) studies computational
complexity in terms of restrictions of languages and computational
principles, providing results that do not depend on specific machine
models. Based on static analysis, it helps predict and control
resources consumed by programs, and can offer reusable and tunable
ideas and techniques for compilers.  ICC mainly focuses on syntactic
\cite{Cob65,BC92}, type \cite{GirardLL,BT-DLAL} and Data Flow
\cite{JonesSCP,Hofmann99,KJmwp,RCG-ToCL} restrictions to provide
bounds on programs' complexity.  The present work was mainly inspired
by the way ICC community uses different concepts to perform Data Flow
Analysis, e.g. ``Size-change Graphs'' \cite{JonesSCP} or ``Resource
Control Graphs''\cite{RCG-ToCL} which track data values' behavior and
use a matrix notation inspired by \cite{AbelAltenkirch}, or
``mwp-polynomials'' \cite{KJmwp} to provide bounds on data size.

For our analysis, we focus on dependencies between variables to detect
invariance. Dependency graphs \cite{Kuck:1981:DGC:567532.567555} can
have different types of arcs representing different kind of
dependencies. Here we will use a kind of Dependence Graph Abstraction
\cite{Cocke:1970:GCS:390013.808480} that can be used to find local and
global quasi-invariants. Based on these techniques, we developed an
analysis pass and we will implement the corresponding transformation
in LLVM.



We propose a tool which is notably able to give enough information to
easily peel and hoist an inner loop, thus decreasing the complexity of
a program from $n^2$ to $n$.

\subsection{State of the art on Quasi-Invariant detection in loop}

Invariants are basically detected using \autoref{Fig_basic}.
\begin{algorithm}[b]
    {\footnotesize
    {
    \KwData{List of Statements in the Loop}
    \KwResult{List of Loop-invariants \texttt{LI}}
    Initialization\;
    \While{search until there is no new invariant…}{
        \For{each statement \texttt{s}}{
            \If{each variable in \texttt{s}\\ has no definition in the
                loop \textbf{or}\\ has exactly one loop-invariant
                definition \textbf{or}\\ is \emph{constant}}{
                Add \texttt{s} to \texttt{LI};
            }
        }
    }
    }
    \caption{Basic invariants detection\label{Fig_basic}}
    }
\end{algorithm}

A \emph{dependency graph} around variables is needed to provide
relations between statements. For quasi-invariance, we need to couple
dependence and \emph{dominance} informations. In \cite{LQICM2000}, the
authors define a \emph{variable dependency graph} (VDG) and detect a
loop quasi-invariant variable \texttt{x} if, among all paths ending at
\texttt{x}, no path contain a node included in a circular path. Then
they deduce an \emph{invariant length} which corresponds to the length
of the longest path ending in \texttt{x}. In the present paper, this
\emph{length} is called \emph{invariance degree}.

\subsection{Contributions}

To the authors' knowledge, this is the first application of ICC
techniques on a mainstream compiler. One interest is that our tool
potentially applies to programs written in any programming language
managed by LLVM\@. Moreover, this work should be considered as a first
step of a larger project that will make ICC techniques more accessible
to programmers.

On a more technical side, our tool aims at improving on currently
implemented loop invariant detection and optimization techniques. The
main LLVM tool for this purpose, Loop Invariant Code Motion (LICM),
does not detect quasi-invariant of degree more than $3$ (and not all
of those of degree $2$). More importantly, LICM will not detect
quasi-invariant blocks of code (what we call \emph{chunk}), such as
whole loops. Our tool, on the other hand, detects quasi-invariants of
arbitrary degree and is able to deal with chunks. For instance the
optimization shown in \autoref{example3} is not performed by LLVM nor
in GCC even at their maximum optimization level.

%% file: part1.tex
\section{In theory}

In this section, we redefine our own types of relations between
variables to build a new dependency graph and apply a composition
inspired by the graph composition of \emph{Size-Change
  Termination}~\cite{JonesSCP}.

\subsection{Relations and Data Flow Graph}


We work with a simple imperative \texttt{WHILE}-language (the grammar is shown in \autoref{fig:grammar}), with
semantics similar to \texttt{C}.

    \begin{figure}[h]
        \centering
        \footnotesize
        \begin{tabular}{llcl}
            (Variables) & \textit{X} & ::= &
            \texttt{$X_1$}\ $|$\ \texttt{$X_2$}\ $|$\ \texttt{$X_3$}\
            $|$\
            …\ $|$\ \texttt{$X_n$}\\
            (Expression) & \textit{exp} & ::= & \texttt{$X$}\ $|$\
            \texttt{op(\emph{exp},…,\emph{exp})}\\
            (Command) & \textit{com} & ::= & \textit{X}=\textit{exp}\ $|$\
            \textit{com;com}\ $|$\ \texttt{skip}\ $|$\ \\
            &&&\texttt{while}\ \textit{exp}\ \texttt{do}\
            \textit{com}\ \texttt{od}\ $|$\
            \\
            &&&\texttt{if}\ \textit{exp}\ \texttt{then}\ \textit{com}\
            \texttt{fi}\ $|$\
            \\
            &&&\texttt{use(\texttt{$X_1$},…,\texttt{$X_n$})}\\
        \end{tabular}
        \caption{Grammar\label{fig:grammar}}
    \end{figure}

A \texttt{WHILE} program is thus a sequence of statements, each
statement being either an \emph{assignment}, a \emph{conditional}, a
\emph{while} loop, a \emph{function call} or a \emph{skip}. The
\emph{use} command represents any command which does not modify its
variables but use them and should not be moved around carelessly
(typically, a \texttt{printf}). \emph{Statements} are abstracted into
\emph{commands}. A \emph{command} can be a statement or a sequence of
commands. We also call a sequence of commands a \emph{chunk}.

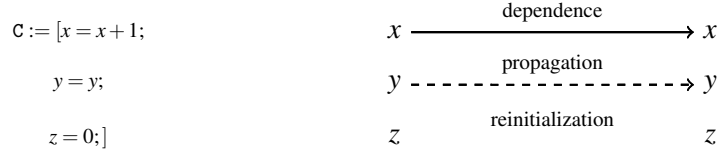
\begin{figure}[h]
    \centering
    \begin{tikzpicture}[scale=0.7]
        \SetVertexMath
        \tikzset{VertexStyle/.style = {shape=circle,
                           minimum size=6pt,inner sep=0pt}
        }
            \node (x0) at (0,0) {$x$};
            \node (x1) at (0,-1) {$y$};
            \node (x2) at (0,-2) {$z$};
        %
        \tikzset{VertexStyle/.style = {shape=circle,
                           minimum size=6pt, inner sep=0pt}
        }
            \node (y0) at (6,0) {$x$};
            \node (y1) at (6,-1) {$y$};
            \node (y2) at (6,-2) {$z$};

        \tikzset{EdgeStyle/.style={->,font=\scriptsize,above,sloped,midway}}
        \Edge[label = dependence](x0)(y0)
        \Edge(x0)(y0)
        \tikzset{
            EdgeStyle/.style={->,font=\scriptsize,above,sloped,midway,dashed}
        }
        \Edge[label = propagation](x1)(y1)
        \Edge(x1)(y1)
        \tikzset{
            EdgeStyle/.style={color=white,->,font=\scriptsize,above,sloped,midway}
        }
        \Edge[label = reinitialization](x2)(y2)
        \Edge(x2)(y2)

        {\scriptsize
        \node[] at (-6,0) {$\comm{C}:= [$\Code{$x=x+1;$}};
        \node[] at (-6,-1) {\Code{$y=y;$}};
        \node[] at (-6,-2) {\Code{$z=0;]$}};
        }

    \end{tikzpicture}
    \caption{Types of dependence\label{Fig_threecases}}
    \label{fig:dependences}
\end{figure}

We start by giving an informal but intuitive definition of the notion
of \emph{Data Flow Graph} (DFG). A DFG represents dependencies between
variables as a bipartite graph as in \autoref{Fig_threecases}. Each
different types of arrow represents different types of dependencies.

Each variable is shown twice: the occurrence on the left represents
the variable before the execution of the command while the occurrence
on the right represents the variable after the execution.
Dependencies are then represented by two types of arrows from
variables on the left to variables on the right: plain arrows for
\emph{direct dependency}, dashed arrows for \emph{propagation}.
\emph{Reinitialisation} of a variable $z$ then corresponds to the
absence of arrows ending on the right occurrence of $z$.
\autoref{Fig_threecases} illustrates these types of dependencies; let
us stress here that the DFG would be the same if the assignment $y=y;$
were to be removed\footnote{Note that $y=y;$ does not create a direct
dependence} from $\comm{C}$ since the value of $y$ is still
propagated.

More formally, a DFG of a command $\comm{C}$ is a triple
$(V,\mathcal{R}_{\mathrm{dep}},\mathcal{R}_{\mathrm{prop}})$ with $V$
the variables involved in the command $\comm{C}$ and a pair of two
relations on the set of variables.  These two relations express how
the values of the involved variables \emph{after} the execution of the
command depend on their values \emph{before} the execution. There is a
\emph{direct} dependence between variables appearing in an expression
and the variable on the left-hand side of the assignment.  For
instance \texttt{$x$} directly depends on $y$ and $z$ in the statement
\Code{$x=y+z;$}.  When variables are unchanged by the command we call
it \emph{propagation}. Propagation only happens when a variable is not
affected by the command, not when it is copied from another variable.
If the variable is set to a constant, we call this a
\emph{reinitialization}.

More technically, we will work with an alternative representation in
terms of matrices. While less intuitive, this allows for more natural
compositions, based on standard linear algebra operations. Before
providing the formal definition, let us introduce the semi-ring
$\{\emptyset, 0, 1\}$: the addition $\oplus$ and multiplication
$\otimes$ are defined in \autoref{Fig_semiring}. Let us remark that,
identifying $\emptyset$ as $-\infty$, this is a sub-semi-ring of the
standard \emph{tropical semi-ring}, with $\oplus$ and $\otimes$
interpreted as $\max$ and $+$ respectively.

\begin{figure}[h]
    \centering
	\hfill
         $\begin{array}{c|ccc}
                \oplus & \emptyset & 0 & 1 \\
                \hline
                \emptyset & \emptyset & 0 & 1 \\
                0 & 0 & 0 & 1 \\
                1 & 1 & 1 & 1 \\
         \end{array}$
	\hfill
         $\begin{array}{c|ccc}
                \otimes & \emptyset & 0 & 1 \\
                \hline
                \emptyset & \emptyset & \emptyset & \emptyset \\
                0 & \emptyset & 0 & 1 \\
                1 & \emptyset & 1 & 1 \\
         \end{array}$
	\hfill ~\\
\caption{Addition and Multiplication in the semi-ring $\{\emptyset, 0, 1\}$.\label{Fig_semiring}}
\end{figure}

\begin{mydef}
  A \emph{Data Flow Graph} for a command $\comm{C}$ is a $n\times n$
  matrix over the semi-ring $\{\emptyset, 0, 1\}$ where $n$ is the
  number of variables involved in $\comm{C}$.

  We write $\dfg{C}$ the DFG of $\comm{C}$. At line $i$, column $j$,
  we have a $\emptyset$ if the output value of the $j$th variable does
  not depend on the input value of the $i$th; a $0$ in case of
  propagation (unmodified variable); and a $1$ for any other kind of
  dependence.
\end{mydef}

\begin{mydef}
Let $\comm{C}$ be a command. We define $\In{\comm{C}}$ (resp.
$\Out{\comm{C}}$) as the set of variables \emph{used} (resp.
\emph{modified}) by $\comm{C}$.
\end{mydef}

Note that $\In{\comm{C}}$ and $\Out{\comm{C}}$ are exactly the set of
variables that are at either ends of the ``dependence'' arrows.

\subsection{Constructing DFGs}

We now describe how the DFG of a command can be computed by induction
on the structure of the command. Base cases (skip, use and assignment)
are done in the obvious way, generalising slightly the definitions of
DFGs shown in \autoref{Fig_threecases}.

\subsubsection{Composition and Multipath}

We now turn to the definition of the DFG for a (sequential)
\emph{composition} of commands. This abstraction allows us to see a
block of statements as one command with its own DFG.

\begin{mydef}
    Let $\comm{C}$ be a sequence of commands
    $[\comm{C_{1}};\comm{C_{2}};\dots;\comm{C_{n}}]$. Then $\dfg{C}$
    is defined as the matrix product
    $\dfg{C_{1}}\dfg{C_{2}}\dots\dfg{C_{n}}$.
\end{mydef}

Following the usual product of matrices, the product of two matrices
$A, B$ is defined here as the matrix $C$ with coefficients:
$C_{i,j}=\bigoplus_{k=1}^{n}(A_{i,k}\otimes B_{k,j})$.

This operation of matrix multiplication corresponds to the computation
of \emph{multipaths} \cite{JonesSCP} in the graph representation of
DFGs. We illustrate this intuitive construction on an example in
\autoref{fig:composition}.

\begin{figure}[h]

    \centering

    \begin{subfigure}{0.4\linewidth}
        \centering
        \begin{tikzpicture}[scale=0.35]
            \SetVertexMath
            \tikzset{VertexStyle/.style = {shape=circle,
                minimum size=6pt,inner sep=0pt}
            }
            \node (x0) at (-2,0) {$w$};
            \node (x1) at (-2,-1) {$x$};
            \node (x2) at (-2,-2) {$y$};
            \node (x3) at (-2,-3) {$z$};
            \node (z0) at (4,0) {$w$};
            \node (z1) at (4,-1) {$x$};
            \node (z2) at (4,-2) {$y$};
            \node (z3) at (4,-3) {$z$};
            \node (y0) at (10,0) {$w$};
            \node (y1) at (10,-1) {$x$};
            \node (y2) at (10,-2) {$y$};
            \node (y3) at (10,-3) {$z$};
            \tikzset{EdgeStyle/.style={->,font=\scriptsize,right,sloped,midway}}
            \Edge[](x0)(z0)
            \Edge[](x1)(z0)
            \Edge[](x2)(z3)
            \Edge[](z2)(y1)
            \Edge[](z3)(y3)

            \tikzset{EdgeStyle/.style={->,font=\scriptsize,above,sloped,midway,dashed}}
            \Edge[](x1)(z1)
            \Edge[](x2)(z2)
            \Edge[](z0)(y0)
            \Edge[](z2)(y2)

            {\scriptsize
                \node[] at (1,1)
                {$\comm{C_{1}}$};
                \node[] at (7,1)
                {$\comm{C_{2}}$};
            }

            {\tiny
                \node [] at (0.8,-5) {
                    $\begin{bmatrix}
                        1 & \emptyset & \emptyset & \emptyset \\
                        1 & 0 & \emptyset & \emptyset \\
                        \emptyset & \emptyset & 0 & 1 \\
                        \emptyset & \emptyset & \emptyset & \emptyset\\
                    \end{bmatrix}
                    $
                };

                \node [] at (7,-5) {
                    $\begin{bmatrix}
                        0 & \emptyset & \emptyset & \emptyset \\
                        \emptyset & \emptyset & \emptyset & \emptyset \\
                        \emptyset & 1 & 0 & \emptyset \\
                        \emptyset & \emptyset & \emptyset & 1\\
                    \end{bmatrix}
                    $
                };
            }

        \end{tikzpicture}
    \end{subfigure}
    \begin{subfigure}{0.4\linewidth}
        \centering
        \begin{tikzpicture}[scale=0.35]
            \SetVertexMath
            \tikzset{VertexStyle/.style = {shape=circle,
                minimum size=6pt,inner sep=0pt}
            }
            \node (x0) at (0,0) {$w$};
            \node (x1) at (0,-1) {$x$};
            \node (x2) at (0,-2) {$y$};
            \node (x3) at (0,-3) {$z$};
            \node (y0) at (6,0) {$w$};
            \node (y1) at (6,-1) {$x$};
            \node (y2) at (6,-2) {$y$};
            \node (y3) at (6,-3) {$z$};
            \tikzset{EdgeStyle/.style={->,font=\scriptsize,right,sloped,midway}}
            \Edge[](x0)(y0)
            \Edge[](x1)(y0)
            \Edge[](x2)(y3)
            \Edge[](x2)(y1)

            Propagation
            \tikzset{EdgeStyle/.style={->,font=\scriptsize,above,sloped,midway,dashed}}
            \Edge[](x2)(y2)

            {\scriptsize
                \node[] at (3,1) {$[\comm{C_{1}};\comm{C_{2}}]$};
            }

            {\tiny
                \node [] at (3,-5) {
                    $\begin{bmatrix}
                        1 & \emptyset & \emptyset & \emptyset \\
                        1 & \emptyset & \emptyset & \emptyset \\
                        \emptyset & 1 & 0 & 1 \\
                        \emptyset & \emptyset & \emptyset & \emptyset\\
                    \end{bmatrix}
                    $
                };
            }
        \end{tikzpicture}
    \label{fig:composition}
    \end{subfigure}
        \caption{
            DFG of Composition.\\
            {\scriptsize
            Here
    $\comm{C_{1}}:=[w=w+x;z=y+2;]$ and
    $\comm{C_{2}}:=[x=y;z=z*2;]$
}
        }\label{fig:composition}
\end{figure}

\subsubsection{Condition}

We now explain how to compute the DFG of a command
$\comm{C}:=\comm{if\ E\ then\ C_{1};}$, from the DFG of the command
$\comm{C_{1}}$.

Firstly, we notice that in $\comm{C}$, all modified variables in
$\comm{C_{1}}$, i.e. in $\Out{\comm{C_{1}}}$, will depend on the
variables used in $\comm{E}$. Let us denote by
$\dfg{C_{1}}^{[\comm{E}]}$ the corresponding DFG, i.e. the matrix
$\dfg{C_{1}}\oplus (E^{t} O)$, where $E$ (resp. $O$) is the vector
representing variables in\footnote{I.e. the vector with a coefficient
equal to $1$ for the variables in $\Var(\comm{E})$, and $\emptyset$
for all others variables.} $\Var(\comm{E})$ (resp. in $\Out{C_{1}}$),
and $(\cdot)^{t}$ denotes the transpose.

Secondly, we need to take into account that the command $\comm{C_{1}}$
may be skipped. In that case, the overall command $\comm{C}$ should
act as an empty command, i.e. be represented by the identity matrix
$\mathrm{Id}$ (diagonal elements are equal to $0$, all other are equal
to $\emptyset$).

Finally, the DFG of a conditional will be computed by summing these
two possibilities, as in \autoref{fig:condition}.

\begin{mydef}
Let $\comm{C}$ be a command of the form $\comm{if\ E\ then\ C_{1};}$.
Then $\dfg{C}=\dfg{C_{1}}^{[\comm{E}]}\oplus\mathrm{Id}$.
\end{mydef}

\begin{figure}[h]
    \centering

    \begin{subfigure}{0.4\linewidth}
        \centering
        \begin{tikzpicture}[scale=0.35]
            \SetVertexMath
            \tikzset{VertexStyle/.style = {shape=circle,
                minimum size=6pt,inner sep=0pt}
            }
            \node (x0) at (7,0) {$w$};
            \node (x1) at (7,-1) {$x$};
            \node (x2) at (7,-2) {$y$};
            \node (x3) at (7,-3) {$z$};
            \node (y0) at (13,0) {$w$};
            \node (y1) at (13,-1) {$x$};
            \node (y2) at (13,-2) {$y$};
            \node (y3) at (13,-3) {$z$};

            \tikzset{EdgeStyle/.style={->,font=\scriptsize,right,sloped,midway}}
            \Edge[](x0)(y0)
            \Edge[](x1)(y0)
            \Edge[](x2)(y3)

            \tikzset{EdgeStyle/.style={->,font=\scriptsize,above,sloped,midway,dashed}}
            \Edge[](x1)(y1)

            {\scriptsize
                \node[] at (2,1)
                {$\comm{E}$};
                \node[] at (4,1)
                {$\comm{O}$};
                \node[] at (10,1)
                {$\comm{C_{1}}$};
            }

            {\tiny
                \node [] at (9.8,-5) {
                    $\begin{bmatrix}
                        1 & \emptyset & \emptyset & \emptyset \\
                        1 & 0 & \emptyset & \emptyset \\
                        \emptyset & \emptyset & 0 & 1 \\
                        \emptyset & \emptyset & \emptyset & \emptyset\\
                    \end{bmatrix}
                    $
                };

                \node [] at (1.8,-5) {
                    $\begin{bmatrix}
                        \emptyset \\
                        \emptyset \\
                        \emptyset \\
                        1 \\
                    \end{bmatrix}
                    $
                };

                \node [] at (3.8,-5) {
                    $\begin{bmatrix}
                        1 \\
                        \emptyset \\
                        1 \\
                        1 \\
                    \end{bmatrix}
                    $
                };
            }

        \end{tikzpicture}
    \end{subfigure}
    \begin{subfigure}{0.4\linewidth}
        \centering
        \begin{tikzpicture}[scale=0.35]
            \SetVertexMath
            \tikzset{VertexStyle/.style = {shape=circle,
                minimum size=6pt,inner sep=0pt}
            }
            \node (x0) at (0,0) {$w$};
            \node (x1) at (0,-1) {$x$};
            \node (x2) at (0,-2) {$y$};
            \node (x3) at (0,-3) {$z$};
            \node (y0) at (6,0) {$w$};
            \node (y1) at (6,-1) {$x$};
            \node (y2) at (6,-2) {$y$};
            \node (y3) at (6,-3) {$z$};
            \tikzset{EdgeStyle/.style={->,font=\scriptsize,right,sloped,midway}}
            \Edge[](x0)(y0)
            \Edge[](x1)(y0)
            \Edge[](x2)(y3)
            \Edge[](x3)(y0)
            \Edge[](x3)(y3)
            \Edge[](x3)(y2)

            \tikzset{EdgeStyle/.style={->,font=\scriptsize,above,sloped,midway,dashed}}
            \Edge[](x1)(y1)
            \Edge[](x2)(y2)

            {\scriptsize
                \node[] at (3,1) {$[\comm{if\ E\ then\ C_{1}}]$};
            }

            {\tiny
                \node [] at (2.8,-5) {
                    $ \begin{bmatrix}
                        1 & \emptyset & \emptyset & \emptyset \\
                        1 & 0 & \emptyset & \emptyset \\
                        \emptyset & \emptyset & \emptyset & 1 \\
                        1 & \emptyset & 1 & 1\\
                    \end{bmatrix}
                    $
                };
            }
        \end{tikzpicture}
    \end{subfigure}
    \caption{DFG of Conditional.\\
            {\scriptsize
Here
$\comm{E}:=$\texttt{$z \geq 0$} and
$\comm{C_{1}}:=$\texttt{$[w=w+x;z=y+2;y=0;];$}
}
}
    \label{fig:condition}
\end{figure}
\vspace{-0.2cm}

\subsubsection{While Loop}

Finally, let us define the DFG of a command $\comm{C}$ of the form
$\comm{C}:=\comm{while\ E\ do\ C_{1};}$. This definition splits into
two steps. First, we define a matrix $\dfg{C_{1}^{\ast}}$ representing
iterations of the command $\comm{C_{1}}$; then we deal with the
condition of the loop in the same way we interpreted the conditional
above.

When considering iterations of $\comm{C_{1}}$, the first occurrence of
$\comm{C_{1}}$ will influence the second one and so on. Computing the
DFG of $\comm{C_{1}^{n}}$, the $n$-th iteration of $\comm{C_{1}}$, is
just computing the power of the corresponding matrix, i.e.
$\dfg{C_{1}^{n}}=\dfg{C_{1}}^{n}$. But since the number of iteration
cannot be decided \emph{a priori}, we need to add all possible values
of $n$. The following expression then expresses the DFG of the
(informal) command $\comm{C_{1}^{\ast}}$ corresponding to "iterating
$\comm{C_{1}}$ a finite (but arbitrary) number of times":
\[\dfg{C_{1}^{\ast}}=\limit_{k \rightarrow \infty}\bigoplus_{i=1}^{k}
\dfg{C_{1}}^i\]
To ease notations, we note $\dfg{C_{1}^{(k)}}$
the partial summations $\sum_{i=1}^{k}\dfg{C_{1}}^{i}$.


Since the set of all relations is finite and the
sequence $(\dfg{C_{1}^{(k)}})_{k\geqslant 0}$ is monotonous, this
sequence is eventually constant. I.e., there exists a natural
number $N$ such that $\dfg{C_{1}^{(k)}}=\dfg{C_{1}^{(N)}}$ for all
$k\geqslant N$. One can obtain the following bound on
the value of $N$.

\begin{mylemma}
    Consider a command $\comm{C}$ and define $K=\min(i,o)$, where $i$
    (resp. $o$) denotes the number of variables in $\In{\comm{C}}$
    (resp. $\Out{\comm{C}}$).  Then, the sequence
    $(\dfg{C^{(k)}})_{k\geqslant K}$ is constant.
\end{mylemma}

\begin{figure}
    \centering
    \begin{subfigure}{0.3\linewidth}
        \centering
        \begin{tikzpicture}[scale=0.30]
            \SetVertexMath
            \tikzset{VertexStyle/.style = {shape=circle,
                minimum size=6pt,inner sep=0pt}
            }
            \node (x0) at (0,0) {$w$};
            \node (x1) at (0,-1) {$x$};
            \node (x2) at (0,-2) {$y$};
            \node (x3) at (0,-3) {$z$};
            \node (y0) at (6,0) {$w$};
            \node (y1) at (6,-1) {$x$};
            \node (y2) at (6,-2) {$y$};
            \node (y3) at (6,-3) {$z$};

            \tikzset{EdgeStyle/.style={->,font=\scriptsize,right,sloped,midway}}
            \Edge[](x0)(y0)
            \Edge[](x1)(y0)
            \Edge[](x2)(y3)
            \Edge[](x2)(y1)

            \tikzset{EdgeStyle/.style={->,font=\scriptsize,above,sloped,midway,dashed}}
            \Edge[](x2)(y2)

            {\scriptsize
                \node[] at (3,1)
                {$\comm{C_{3}}$};
            }

            {\tiny
                \node [] at (3,-5.5) {
                    $ \begin{bmatrix}
                        1 & \emptyset & \emptyset & \emptyset \\
                        1 & \emptyset & \emptyset & \emptyset \\
                        \emptyset & 1 & 0 & 1 \\
                        \emptyset & \emptyset & \emptyset & \emptyset\\
                    \end{bmatrix}
                    $
                };

            }

        \end{tikzpicture}
    \end{subfigure}
    ~
    \begin{subfigure}{0.3\linewidth}
        \centering
        \begin{tikzpicture}[scale=0.30]
            \SetVertexMath
            \tikzset{VertexStyle/.style = {shape=circle,
                minimum size=6pt,inner sep=0pt}
            }
            \node (x0) at (0,0) {$w$};
            \node (x1) at (0,-1) {$x$};
            \node (x2) at (0,-2) {$y$};
            \node (x3) at (0,-3) {$z$};
            \node (y0) at (6,0) {$w$};
            \node (y1) at (6,-1) {$x$};
            \node (y2) at (6,-2) {$y$};
            \node (y3) at (6,-3) {$z$};

            \tikzset{EdgeStyle/.style={->,font=\scriptsize,right,sloped,midway}}
            \Edge[](x0)(y0)
            \Edge[](x1)(y0)
            \Edge[](x2)(y3)
            \Edge[](x2)(y1)
            \Edge[](x2)(y0)

            \tikzset{EdgeStyle/.style={->,font=\scriptsize,above,sloped,midway,dashed}}
            \Edge[](x2)(y2)

            {\scriptsize
                \node[] at (3,1)
                {$\comm{C_{3}^{(2)}}$};
            }

            {\tiny
                \node [] at (3,-5.5) {
                    $ \begin{bmatrix}
                        1 & \emptyset & \emptyset & \emptyset \\
                        1 & \emptyset & \emptyset & \emptyset \\
                        1 & 1 & 0 & 1 \\
                        \emptyset & \emptyset & \emptyset & \emptyset\\
                    \end{bmatrix}
                    $
                };

            }

        \end{tikzpicture}
    \end{subfigure}
    ~
    \begin{subfigure}{0.3\linewidth}
        \centering
        \begin{tikzpicture}[scale=0.30]
            \SetVertexMath
            \tikzset{VertexStyle/.style = {shape=circle,
                minimum size=6pt,inner sep=0pt}
            }
            \node (x0) at (0,0) {$w$};
            \node (x1) at (0,-1) {$x$};
            \node (x2) at (0,-2) {$y$};
            \node (x3) at (0,-3) {$z$};
            \node (y0) at (6,0) {$w$};
            \node (y1) at (6,-1) {$x$};
            \node (y2) at (6,-2) {$y$};
            \node (y3) at (6,-3) {$z$};

            \tikzset{EdgeStyle/.style={->,font=\scriptsize,right,sloped,midway}}
            \Edge[](x0)(y0)
            \Edge[](x1)(y0)
            \Edge[](x2)(y3)
            \Edge[](x2)(y1)
            \Edge[](x2)(y0)

            \tikzset{EdgeStyle/.style={->,font=\scriptsize,above,sloped,midway,dashed}}
            \Edge[](x2)(y2)

            {\scriptsize
                \node[] at (3,1)
                {$\comm{C_{3}^{(3)}}$};
            }

            {\tiny
                \node [] at (3,-5.5) {
                    $\begin{bmatrix}
                        1 & \emptyset & \emptyset & \emptyset \\
                        1 & \emptyset & \emptyset & \emptyset \\
                        1 & 1 & 0 & 1 \\
                        \emptyset & \emptyset & \emptyset & \emptyset\\
                    \end{bmatrix}
                    $
                };

            }

        \end{tikzpicture}
    \end{subfigure}
    \caption{
        Finding fix point of dependence (simple example)\\
        {\scriptsize
Here $\comm{C_{3}}:=$\texttt{$[w=w+x;z=y+2;x=y;z=z*2];$}\\
}
    }\label{Fig_loop}
\end{figure}
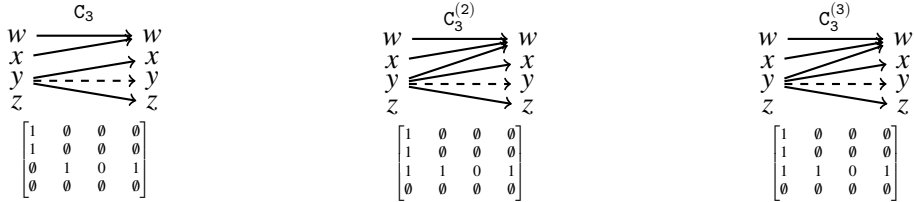

\autoref{Fig_loop} illustrates the computation of $\cdot^{\ast}$.
The second step then consists in dealing with the loop condition,
using the same constructions as for conditionals.

\begin{mydef}
Let $\comm{C}$ be a command of the form $\comm{while\ E\ do\ C_{1};}$.
Then $\dfg{C}=\dfg{C_{1}^{\ast}}^{[\comm{E}]}$.
\end{mydef}

\subsection{Independence}

Our purpose is to move commands around: exchange them, but more
importantly to pull them out of loops when possible.  We allow these
moves only when semantics are preserved: to ensure this is the case,
we describe a notion of independence.

\vspace{-0.2cm}
\begin{mydef}
    If $\Out{\comm{C_{1}}} \cap \In{\comm{C_{2}}} = \emptyset$ then
    $\comm{C_{2}}$ is \emph{independent} from $\comm{C_{1}}$. This is
    denoted $\comm{C_{1}}\prec\comm{C_{2}}$.
\end{mydef}

It is important to notice that this notion is not symmetric. As an
example, let us consider \autoref{compoind}:
\begin{figure}[h]
    \centering

    \begin{subfigure}{0.55\linewidth}
        \centering
        \begin{tikzpicture}[scale=0.35]
            \SetVertexMath
            \tikzset{VertexStyle/.style = {shape=circle,
                       minimum size=6pt,inner sep=0pt}
                }
            \node (x0) at (-4,0) {$w$};
            \node (x1) at (-4,-1) {$x$};
            \node (x2) at (-4,-2) {$y$};
            \node (x3) at (-4,-3) {$z$};
            \node (z0) at (2,0) {$w$};
            \node (z1) at (2,-1) {$x$};
            \node (z2) at (2,-2) {$y$};
            \node (z3) at (2,-3) {$z$};
            \node (y0) at (8,0) {$w$};
            \node (y1) at (8,-1) {$x$};
            \node (y2) at (8,-2) {$y$};
            \node (y3) at (8,-3) {$z$};

            \tikzset{EdgeStyle/.style={->,font=\scriptsize,right,sloped,midway}}
            \Edge[](x0)(z0)
            \Edge[](x1)(z0)
            \Edge[](z2)(y1)
            \Edge[](z3)(y3)

            \tikzset{EdgeStyle/.style={->,font=\scriptsize,above,sloped,midway,dashed}}
            \Edge[](x1)(z1)
            \Edge[](x2)(z2)
            \Edge[](z0)(y0)
            \Edge[](z2)(y2)
            \Edge[](x3)(z3)

            {\scriptsize
            \node[] at (-1,1)
            {$\comm{C_{1}}$};
            \node[] at (5,1)
            {$\comm{C_{2}}$};
            }

            {\tiny
            \node [] at (-1.2,-5) {
                $ \begin{bmatrix}
                    1 & \emptyset & \emptyset & \emptyset \\
                    1 & 0 & \emptyset & \emptyset \\
                    \emptyset & \emptyset & 0 & \emptyset \\
                    \emptyset & \emptyset & \emptyset & 0\\
                \end{bmatrix}
            $
        };

            \node [] at (4.9,-5) {
                $ \begin{bmatrix}
                    0 & \emptyset & \emptyset & \emptyset \\
                    \emptyset & \emptyset & \emptyset & \emptyset \\
                    \emptyset & 1 & 0 & \emptyset \\
                    \emptyset & \emptyset & \emptyset & 1\\
                \end{bmatrix}
            $
        };
            }

        \end{tikzpicture}
    \end{subfigure}
~
\begin{subfigure}{0.45\linewidth}
    \centering
    \begin{tikzpicture}[scale=0.35]
        \SetVertexMath
        \tikzset{VertexStyle/.style = {shape=circle,
                       minimum size=6pt,inner sep=0pt}
                }
            \node (x0) at (0,0) {$w$};
            \node (x1) at (0,-1) {$x$};
            \node (x2) at (0,-2) {$y$};
            \node (x3) at (0,-3) {$z$};
            \node (y0) at (6,0) {$w$};
            \node (y1) at (6,-1) {$x$};
            \node (y2) at (6,-2) {$y$};
            \node (y3) at (6,-3) {$z$};
        \tikzset{EdgeStyle/.style={->,font=\scriptsize,right,sloped,midway}}
        \Edge[](x0)(y0)
        \Edge[](x1)(y0)
        \Edge[](x3)(y3)
        \Edge[](x2)(y1)

        \tikzset{EdgeStyle/.style={->,font=\scriptsize,above,sloped,midway,dashed}}
        \Edge[](x2)(y2)

        {\scriptsize
        \node[] at (3,1) {$[\comm{C_{1}};\comm{C_{2}}]$};
        }

        {\tiny
        \node [] at (2.9,-5) {
            $\begin{bmatrix}
                1 & \emptyset & \emptyset & \emptyset \\
                1 & \emptyset & \emptyset & \emptyset \\
                \emptyset & 1 & 0 & \emptyset \\
                \emptyset & \emptyset & \emptyset & 1\\
            \end{bmatrix}
            $
        };
        }
    \end{tikzpicture}
\end{subfigure}
    \caption{
        Composition of independent chunks of commands\\
         {\scriptsize
         Here
$\comm{C_{1}}:=$\texttt{$[w=w+x;]$} and
$\comm{C_{2}}:=$\texttt{$[x=y;z=z*2];$}
}
    }
    \label{compoind}
\end{figure}
Here, $\comm{C_{2}}$ is independent from $\comm{C_{1}}$ but
the inverse is not true.

A particular case is \emph{self-independence}, i.e. independence of a
command w.r.t. itself. In that case, we can find non-trivial program
transformations preserving the semantics. We denote by
$\llbracket\comm{C}\rrbracket\equiv\llbracket\comm{D}\rrbracket$ the
relation "$\comm{C}$ and $\comm{D}$ have the same semantics".

\begin{mylemma}[Specialization for \texttt{while}]
    If $\comm{C_{1}}$ is self-independent and $\Var(\comm{E}) \cap
    \Out{C_{1}} = \emptyset$: 
    \[\llbracket\comm{while\ E\ do\ C_{1}}\rrbracket \equiv
    \llbracket\comm{if\ E\ then\ C_{1};While\ E\ do\ skip}\rrbracket\]
\end{mylemma}

Remark that we need to keep the loop $\comm{While}$ with a skip
statement inside because we need to consider an infinite loop if
$\comm{E}$ is always true to keep the semantic equivalent.

In general, we will consider mutual independence.

\begin{mydef}
    If $\comm{C_{2}}\prec \comm{C_{1}}$ and $\comm{C_{1}}\prec
    \comm{C_{2}}$, we say that $\comm{C_{2}}$ and $\comm{C_{1}}$ are
    mutually independents, and write $\comm{C_{1}} \asymp
    \comm{C_{2}}$.
\end{mydef}

While independence in one direction only, such as in the example
above, does not imply that  $\comm{C_{1}}; \comm{C_{2}}$ and
$\comm{C_{2}};\comm{C_{1}}$ have the same semantics, mutual
independence allows to perform program transformation that do not
impact the semantics.

\begin{mylemma}[Swapping commands]
  If $\comm{C_{1}} \asymp \comm{C_{2}}$, then
  $\llbracket\comm{C_{1}};\comm{C_{2}}\rrbracket \equiv
  \llbracket\comm{C_{2}};\comm{C_{1}}\rrbracket$
\end{mylemma}

\begin{mylemma}[Moving out of \texttt{while} loops]
  If $\comm{C_{1}}$ is self-independent (i.e.
  $\comm{C_{1}} \asymp \comm{C_{1}}$), and if
  $\comm{C_{1}} \asymp \comm{C_{2}}$, then:
    \[
      \llbracket\comm{while\ E\ do\ [C_{1};C_{2}]}\rrbracket \equiv
      \llbracket\comm{if\ E\ then\ C_{1}; while\ E\ do\
        C_{2}}\rrbracket
    \]
\end{mylemma}

Based on those lemmas, we can decide that an entire block of statement
is invariant or quasi-invariant in a loop by computing the DFGs. The
quasi-invariance comes with an \emph{invariance degree} wich is the
number of time the loop needs to be peeled to be able to hoist the
corresponding invariant.  We can then implement program
transformations that reduce the overall complexity while preserving
the semantics.

%% file: part2.tex
\section{In practice}

This section explains how we implemented the pass which computes the
\emph{invariance degree} and gives the main idea of how the
transformation can be performed. In the previous Section, we have seen
that the transformation is possible from and to a \textsc{While}
language; and from a previous 
implementation\footnote{\url{https://github.com/ThomasRuby/LQICM\_On\_C\_Toy\_Parser}}, 
we have shown it can be
done on \texttt{C} Abstract Syntax Trees.

Compilers, and especially LLVM on which we are working, use an
\emph{Intermediate Representation} to handle programs. This is an
assembly-like language that is used during all the stages of the
compilation. Programs (in various different languages) are first
translated into the IR, then several optimisations are performed
(implemented in so-called \emph{passes}), and finally the resulting IR
is translated again in actual assembly language depending on the
machine it will run on. Using a common IR allows to do the same
optimisations on several different source languages and for several
different target architectures.

One important feature of the LLVM IR is the \emph{Single Static
Assignment} form (SSA). A program is in SSA form if each variable is
assigned at most once. In other words, setting a program in SSA form
require a massive $\alpha$-conversion of all the variables to ensure
uniqueness of names. The advantages are obvious since this removes any
name-aliasing problem and ease analysis and transformation.

The main drawback of SSA comes when several different paths in the
Control Flow reach the same point (typically, after a
conditional). Then, the values used after this point may come from any
branch and this cannot be statically decided. For example, if the
original program is \texttt{if (y) then x:=0 else x:=1;C}, it is
relatively easy to turn it into a pseudo-SSA form by $\alpha$-converting
the \texttt{x}: \texttt{if (y) then x$_0$:=0 else x$_1$:=1;C} but we
do not know in \texttt{C} which of \texttt{x$_0$} or \texttt{x$_1$}
should be used.

SSA solves this problem by using \emph{$\varphi$-functions} that, at
runtime, can choose the correct value depending on the path just
taken. That is, the correct SSA form will be \texttt{if (y) then
  x$_0$:=0 else x$_1$:=1; X:=$\varphi$(x$_0$, x$_1$); C}.

While the SSA itself eases the analysis, we do have to take into
account the $\varphi$ functions and handle them correctly.

\subsection{Preliminaries}

First, we want to visit all loops using a bottom-up strategy (the
inner loop first). Then, as for the \emph{Loop Invariant Code Motion}
(LICM) pass, our pass is derived from the basic
\texttt{LoopPass}. Which means that each time a loop is encountered,
our analysis is performed.

At this point, the purpose is to gather the relations of all
instructions in the loop to compose them and provide the final
relation for the current loop. We decided to define a
\texttt{Relation} object by three \texttt{SmallPtrSet} of
\texttt{Value*}, listing the \emph{variables}, the \emph{propagations}
and the \emph{initializations}. Furthermore we represent the
\emph{dependencies} by a \texttt{DenseMap} of \texttt{Value*} to
\texttt{SmallPtrSet<Value*>}. This way of representing our data is not
fixed, it's certainly optimizable, but we think it's sufficient for
our prototype analysis and examples. We will discuss the cost of this
analysis later.

Then a \texttt{Relation} is generated for each command using a
top-down strategy following the dominance tree. The \emph{SSA} form
helps us to gather dependence information on instructions. By visiting
operands of each assignment, it's easy to build our map of
\texttt{Relation}.  With all the current loop's relations gathered, we
compute the compositions, condition corrections and the maximums
relations possible as described previously.  Obviously this method can
be enhanced by an analysis on bounds around conditional and number of
iterations for a loop.  Finally, with those composed relations we
compute an invariance degree for each statement in the loop.

The only chunks considered in the current implementation are the one
consisting of \texttt{while} or \texttt{if-then-else} statements.

\subsection{Invariance degree computation}

In this part, we will describe an algorithm -- using the previous
concepts -- to compute the invariance degree of each quasi-invariant
in a loop. After that, we will be able to peel the loop at once
instead of doing it iteratively.  To simplify and as a recall,
\autoref{example2} shows a basic example of peeled loop.

\label{use}
The invariance degrees are given as comment after each
Quasi-Invariant statements.
So $\texttt{b=y+y}$ is invariant of degree equal to one because
\texttt{y} is invariant, that means it could be hoisted directly in
the \emph{preheader} of the loop. But $\texttt{b}$ is used before, in
$\texttt{b=b+1}$, so it's not the same $\texttt{b}$ at the first
iteration. We need to isolate this case by peeling one time the entire
loop to use the first \texttt{b} computed by the initial
\texttt{b}. If $\texttt{b=y+y}$ is successfully hoisted, then
\texttt{b} is now invariant. So we can remove \texttt{b=b+1} but we
need to do it at least one time after the first iteration to set
\texttt{b} to the new and invariant value. This is why the loop is
peeled two times. The first time, all the statements are executed.
The second time, the first degree invariants are removed.  The main
work is to compute the proper invariance degree for each statement and
composed statements. This can be done statically using the dependency
graph and dominance graph. Here is the algorithm.  Let suppose we have
computed the list of dependencies for all commands in a loop.

\begin{algorithm}
    {\footnotesize
    \KwData{Dependency Graph and Dominance Graph}
    \KwResult{List of invariance degree for each statement}
    Initialize degrees of \texttt{use} to $\infty$ and others to $0$\;
    \For{each statement \texttt{s}}{
        \eIf{the current degree \texttt{cd} $\neq 0$}{
            skip
        }
        {
            Initialize the current degree \texttt{cd} to $\infty$\;
            \eIf{there is no dependence for the current chunk}{
                $\texttt{cd} = 1$\;
            }{
                \For{each dependence compute the degree \texttt{dd} of the
                command}{
                    \eIf{$\texttt{cd}\leq\texttt{dd}$ \textbf{and} the current command
                    dominates this dependence}{
                        $\texttt{cd}=\texttt{dd}+1$
                    }{
                        $\texttt{cd}=\texttt{dd}$
                    }
                }
            }
        }
    }
    \caption{Invariance degree computation.\label{Fig_invariance}}
    }
\end{algorithm}

This algorithm is dynamic. It stores progressively each degree needed
to compute the current one and reuse them.  Note that, for the
initialization part, we are using LLVM methods
(\texttt{canSinkOrHoist}, \texttt{isGuaranteedToExecute} etc…) to
figure out if an instruction is movable or not. These methods provide
the anchors instructions for the current loop.

\subsection{Peeling loop idea}

The transformation will consists in creating as many
\texttt{preheaders} basic blocks before the loop as needed to remove
all quasi-invariants out of the loop. Each \texttt{preheader} will
have the same condition as the \texttt{.cond} block of the loop and
will contain the incrementation of the iteration variable.  The
maximum invariance degree is the number of time we need to peel the
loop. So we can create as many \texttt{preheaders} before the loop.
For each block created, we include every commands with a higher or
equal invariance degree. For instance, the first \texttt{preheader}
block will contain every commands with an invariance degree higher or
equal to $1$, the second one, higher or equal to $2$ etc… and the
final loop will contain every commands with an invariance degree equal
to $\infty$.


%
\begin{figure}
\lstset{
    language=c,
    showstringspaces=false,
    columns=flexible,
    basicstyle={\scriptsize\ttfamily},
    numbers=left,
    numberstyle=\tiny\color{gray},
    stepnumber=1,
    numbersep=8pt,
    firstnumber=0,
    numberfirstline=true,
    commentstyle=\color{blue},
    breaklines=true,
    breakatwhitespace=true,
    emph={use},
    emphstyle={\bf},
    literate={B}{{\color{red}\textbf{b\_1}}}1
    {b}{{\color{blue}\textbf{b}}}1 {S}{{\hspace{1.2em}}}1
}
\begin{tikzpicture}
\node (before) at (0,7) {
		\begin{minipage}[t]{0.35\linewidth}
\begin{lstlisting}
    while(x<100){
        b=b+1; //2
        use(b);
        x=x+1;
        b=y+y; //1
        use(b);
    }
\end{lstlisting}
		\end{minipage}
		};
\node (after) at (5,5) {
	\begin{minipage}[c]{0.35\linewidth}
\begin{lstlisting}
  if (x < 100) //1
  {
    B = b+1;
    use(B);
    x = x+1;
    b = y+y;
    use(b);
  }
  if (x < 100) //2
  {
    B = b+1;
    use(B);
    x = x+1;
    use(b);
  }
  while (x < 100)
  {
    use(B);
    x = x+1;
    use(b);
  }
\end{lstlisting}
	\end{minipage}
	};
\node (afterwest) at (1.8,5) {};
\draw[->] (before.south) .. controls (0,4.8) and (-0.2,5) .. (afterwest) node [sloped,above,near end] {peeling};
\end{tikzpicture}
\caption{Example: Hoisting twice.}\label{example2}
\end{figure}

%% file: conclusion.tex
\section{Conclusion and Future work}

Developers expect that compilers provide certain more or less
``obvious'' optimizations. When peeling is possible, that often means:
either the code was generated; or the developers prefer this form (for
readability reasons) and expect that it will be optimized by the
compiler; or the developers haven't seen the possible optimization
(mainly because of the obfuscation level of a given code).

Our generic pass is able to provide a reusable abstract dependency
graph and the quasi-invariance degrees for further loop optimization
or analysis.

\begin{figure}
\lstset{
    language=c,
showstringspaces=false,
columns=flexible,
basicstyle={\scriptsize\ttfamily},
numbers=left,
numberstyle=\tiny\color{gray},
stepnumber=1,
numbersep=8pt,
firstnumber=0,
numberfirstline=true,
commentstyle=\color{blue},
breaklines=true,
breakatwhitespace=true,
emph={use},
emphstyle={\bf},
literate={B}{{\color{red}\textbf{B}}}1 {S}{{\hspace{1.2em}}}1
}
\begin{tikzpicture}
\node (before) at (0,6) {
		\begin{minipage}[t]{0.35\linewidth}
\begin{lstlisting}
    srand(time(NULL));
    int n=rand()%10000;
    int j=0; 
    while(j<n){
        fact=1;
        i=1;
        while (i<=n) { 
            fact=fact*i; 
            i=i+1;
        }
        j=j+1;
        use(fact);
    }
\end{lstlisting}
		\end{minipage}
		};
\node (after) at (6,5) {
	\begin{minipage}[c]{0.45\linewidth}
\begin{lstlisting}
  srand(time(NULL));
  int n = rand() % 10000;
  int j = 0;
  if (j < n)
  {
    fact = 1;
    i = 1;
    while (i <= n)
    {
      fact = fact * i;
      i = i + 1;
    }
    j = j + 1;
    use(fact);
  }
  while (j < n)
  {
    j = j + 1;
    use(fact);
  }
\end{lstlisting}
	\end{minipage}
	};
\node (afterwest) at (2,3) {};
\draw[->] (before.south) .. controls (0,2.8) and (-0.2,3) .. (afterwest) node [sloped,above,near end] {peeling};
\end{tikzpicture}
\caption{Hoisting inner loop}\label{example3}
\end{figure}

In this example (\autoref{example3}), we compute the same factorial
several times. We can detect it statically, so the compiler has to
optimize it at least in \texttt{-O3}. Our tests showed that is done
neither in LLVM nor in GCC (we also tried \texttt{-fpeel\_loops} with
profiling). The generated assembly shows the factorial computation in
the inner loop.

Moreover, the computation time of this kind of algorithm compiled with
\texttt{clang} in \texttt{-O3} still computes $n$ times the inner loop
so the computation time is quadratic, while hoisting it result in
linear time. For the example shown in \autoref{example3}, our pass
computes the degrees shown in \autoref{example3Degree} (where $-1$
represents a degree of $\infty$, that is an instruction that cannot be
hoisted).

\begin{figure}
\lstset{
    language=llvm,
style=nasm,
showstringspaces=false,
columns=flexible,
basicstyle={\tiny\ttfamily},
numberstyle=\tiny\color{gray},
commentstyle=\color{blue},
breaklines=true,
breakatwhitespace=true,
emph={use},
emphstyle={\bf},
literate={B}{{\color{red}\textbf{B}}}1 {S}{{\hspace{1.2em}}}1
}
\begin{tikzpicture}
\node (before) at (0,6) {
		\begin{minipage}[t]{\linewidth}
\begin{lstlisting}
...
while.cond:
  %i.0 = phi i32 [ 1, %entry ], [ %i.1.lcssa, %while.end ]
  %j.0 = phi i32 [ 0, %entry ], [ %add8, %while.end ]
  %exitcond = icmp ne i32 %j.0, 100
  br i1 %exitcond, label %while.body, label %while.end9

while.body:
  br label %while.cond3

while.cond3:
  %fact.0 = phi i32 [ 1, %while.body ], [ %mul, %while.body6 ]
  %i.1 = phi i32 [ 1, %while.body ], [ %add, %while.body6 ]
  %cmp4 = icmp slt i32 %i.1, %rem
  br i1 %cmp4, label %while.body6, label %while.end

while.body6:
  %mul = mul nsw i32 %fact.0, %i.1
  %add = add nuw nsw i32 %i.1, 1
  br label %while.cond3

while.end:
  %fact.0.lcssa = phi i32 [ %fact.0, %while.cond3 ]
  %i.1.lcssa = phi i32 [ %i.1, %while.cond3 ]
  %call7 = call ... i32 %i.1.lcssa, i32 %fact.0.lcssa)
  %add8 = add nuw nsw i32 %j.0, 1
  br label %while.cond
...
\end{lstlisting}
		\end{minipage}
		};
\end{tikzpicture}
\caption{LLVM Intermediate Representation}\label{example3IR}
\end{figure}

\begin{figure}
\lstset{
    language=llvm,
style=nasm,
showstringspaces=false,
columns=flexible,
basicstyle={\scriptsize\ttfamily},
numberstyle=\tiny\color{gray},
commentstyle=\color{blue},
breaklines=true,
breakatwhitespace=true,
emph={use},
emphstyle={\bf},
literate={B}{{\color{red}\textbf{B}}}1 {S}{{\hspace{1.2em}}}1
}
\begin{tikzpicture}
\node (before) at (0,6) {
		\begin{minipage}[t]{\linewidth}
\begin{lstlisting}
---- MapDeg of while.cond3 ----
%mul = mul nsw i32 %fact.1, %i.1 = -1
%add = add nuw nsw i32 %i.1, 1 = -1
--------------------

---- MapDeg of while.cond ----
%fact.0 = phi i32 [ 1, %while.body ], ... = 1
%i.1 = phi i32 [ 1, %while.body ], ... = 1
inner loop starting with while.cond3: = 1
%fact.0.lcssa = phi i32 [ %fact.0, %while.cond3 ] = -1
%i.1.lcssa = phi i32 [ %i.1, %while.cond3 ] = -1
%call7 = call ... i32 %i.1.lcssa, i32 %fact.0.lcssa) = -1
%add8 = add nuw nsw i32 %j.0, 1 = -1
--------------------
\end{lstlisting}
		\end{minipage}
		};
\end{tikzpicture}
\caption{Invariance Degree}\label{example3Degree}
\end{figure}

To each instruction printed corresponds an invariance degree.  The
assignment instructions are listed by loops, the inner loop (starting
with \texttt{while.cond3}) and the outer loop (starting with
\texttt{while.cond}). The inner loop has its own invariance degree
equal to $1$ (line $9$). Remark that we do consider the \texttt{phi}
initialization instructions of an inner loop. Here \texttt{\%fact.0}
and \texttt{\%i.1} are reinitialized in the inner loop condition
block. So \texttt{phi} instructions are analysed in two different
cases: to compute the relation of the current loop or to give the
initialization of a variable sent to an inner loop. Our analysis only
takes the relevant operand regarding to the current case and do not
consider others.


The code of this pass is available
online\footnote{\url{https://github.com/ThomasRuby/lqicm\_pass}}.  To
provide some real benchmarks on large programs we need to implement
the transformation. We are currently implementing this second pass on
LLVM.


%% file: main.bbl
\begin{thebibliography}{10}
\providecommand{\bibitemdeclare}[2]{}
\providecommand{\surnamestart}{}
\providecommand{\surnameend}{}
\providecommand{\urlprefix}{Available at }
\providecommand{\url}[1]{\texttt{#1}}
\providecommand{\href}[2]{\texttt{#2}}
\providecommand{\urlalt}[2]{\href{#1}{#2}}
\providecommand{\doi}[1]{doi:\urlalt{http://dx.doi.org/#1}{#1}}
\providecommand{\bibinfo}[2]{#2}

\bibitemdeclare{article}{AbelAltenkirch}
\bibitem{AbelAltenkirch}
\bibinfo{author}{A.~\surnamestart Abel\surnameend} \&
  \bibinfo{author}{T.~\surnamestart Altenkirch\surnameend}
  (\bibinfo{year}{2002}): \emph{\bibinfo{title}{A {P}redicative {A}nalysis of
  {S}tructural {R}ecursion}}.
\newblock {\sl \bibinfo{journal}{Journal of Functional Programming}}
  \bibinfo{volume}{12}(\bibinfo{number}{1}), \doi{10.1017/S0956796801004191}.

\bibitemdeclare{article}{BT-DLAL}
\bibitem{BT-DLAL}
\bibinfo{author}{Patrick \surnamestart Baillot\surnameend} \&
  \bibinfo{author}{Kazushige \surnamestart Terui\surnameend}
  (\bibinfo{year}{2009}): \emph{\bibinfo{title}{Light types for polynomial time
  computation in lambda calculus}}.
\newblock {\sl \bibinfo{journal}{Information and {C}omputation}}
  \bibinfo{volume}{207}(\bibinfo{number}{1}), pp. \bibinfo{pages}{41--62},
  \doi{10.1016/j.ic.2008.08.005}.

\bibitemdeclare{article}{BC92}
\bibitem{BC92}
\bibinfo{author}{S.~\surnamestart Bellantoni\surnameend} \&
  \bibinfo{author}{S.~\surnamestart Cook\surnameend} (\bibinfo{year}{1992}):
  \emph{\bibinfo{title}{A new recursion-theoretic characterization of the
  poly-time functions}}.
\newblock {\sl \bibinfo{journal}{Computational Complexity}}
  \bibinfo{volume}{2}, \doi{10.1007/BF01201998}.

\bibitemdeclare{incollection}{Cob65}
\bibitem{Cob65}
\bibinfo{author}{A.~\surnamestart Cobham\surnameend} (\bibinfo{year}{1962}):
  \emph{\bibinfo{title}{The intrinsic computational difficulty of functions}}.
\newblock In \bibinfo{editor}{Y.~\surnamestart Bar-Hillel\surnameend}, editor:
  {\sl \bibinfo{booktitle}{CLMPS}}, \doi{10.2307/2270886}.

\bibitemdeclare{article}{Cocke:1970:GCS:390013.808480}
\bibitem{Cocke:1970:GCS:390013.808480}
\bibinfo{author}{John \surnamestart Cocke\surnameend} (\bibinfo{year}{1970}):
  \emph{\bibinfo{title}{Global Common Subexpression Elimination}}.
\newblock {\sl \bibinfo{journal}{SIGPLAN Not.}}
  \bibinfo{volume}{5}(\bibinfo{number}{7}), \doi{10.1145/390013.808480}.

\bibitemdeclare{article}{GirardLL}
\bibitem{GirardLL}
\bibinfo{author}{J.-Y. \surnamestart Girard\surnameend} (\bibinfo{year}{1987}):
  \emph{\bibinfo{title}{Linear Logic}}.
\newblock {\sl \bibinfo{journal}{Th. Comp. Sci.}} \bibinfo{volume}{50},
  \doi{10.1016/0304-3975(87)90045-4}.

\bibitemdeclare{inproceedings}{Hofmann99}
\bibitem{Hofmann99}
\bibinfo{author}{M.~\surnamestart Hofmann\surnameend} (\bibinfo{year}{1999}):
  \emph{\bibinfo{title}{Linear types and {N}on-{S}ize {I}ncreasing polynomial
  time computation}}.
\newblock In: {\sl \bibinfo{booktitle}{LICS}}, pp. \bibinfo{pages}{464--473},
  \doi{10.1109/LICS.1999.782641}.

\bibitemdeclare{article}{KJmwp}
\bibitem{KJmwp}
\bibinfo{author}{Neil~D. \surnamestart Jones\surnameend} \&
  \bibinfo{author}{Lars \surnamestart Kristiansen\surnameend}
  (\bibinfo{year}{2009}): \emph{\bibinfo{title}{A flow calculus of
  \emph{mwp}-bounds for complexity analysis}}.
\newblock {\sl \bibinfo{journal}{Trans. Comp. Logic}}
  \bibinfo{volume}{10}(\bibinfo{number}{4}), pp. \bibinfo{pages}{28:1--28:41},
  \doi{10.1145/1555746.1555752}.

\bibitemdeclare{unpublished}{LarsDraft}
\bibitem{LarsDraft}
\bibinfo{author}{L.~\surnamestart Kristiansen\surnameend}:
  \emph{\bibinfo{title}{Notes on Code Motion}}.
\newblock \bibinfo{note}{Manuscript}.

\bibitemdeclare{inproceedings}{Kuck:1981:DGC:567532.567555}
\bibitem{Kuck:1981:DGC:567532.567555}
\bibinfo{author}{D.~J. \surnamestart Kuck\surnameend}, \bibinfo{author}{R.~H.
  \surnamestart Kuhn\surnameend}, \bibinfo{author}{D.~A. \surnamestart
  Padua\surnameend}, \bibinfo{author}{B.~\surnamestart Leasure\surnameend} \&
  \bibinfo{author}{M.~\surnamestart Wolfe\surnameend} (\bibinfo{year}{1981}):
  \emph{\bibinfo{title}{Dependence Graphs and Compiler Optimizations}}.
\newblock In: {\sl \bibinfo{booktitle}{POPL}}, \doi{10.1145/567532.567555}.

\bibitemdeclare{inproceedings}{JonesSCP}
\bibitem{JonesSCP}
\bibinfo{author}{C.~S. \surnamestart Lee\surnameend}, \bibinfo{author}{N.~D.
  \surnamestart Jones\surnameend} \& \bibinfo{author}{A.~M. \surnamestart
  Ben-Amram\surnameend} (\bibinfo{year}{2001}): \emph{\bibinfo{title}{The
  {S}ize-{C}hange {P}rinciple for {P}rogram {T}ermination}}.
\newblock In: {\sl \bibinfo{booktitle}{POPL}}, \doi{10.1145/360204.360210}.

\bibitemdeclare{article}{RCG-ToCL}
\bibitem{RCG-ToCL}
\bibinfo{author}{Jean-Yves \surnamestart Moyen\surnameend}
  (\bibinfo{year}{2009}): \emph{\bibinfo{title}{{Resource control graphs}}}.
\newblock {\sl \bibinfo{journal}{ACM Trans. Computational Logic}}
  \bibinfo{volume}{10}, \doi{10.1145/360204.360210}.

\bibitemdeclare{article}{LQICM2000}
\bibitem{LQICM2000}
\bibinfo{author}{Litong \surnamestart Song\surnameend},
  \bibinfo{author}{Yoshihiko \surnamestart Futurama\surnameend},
  \bibinfo{author}{Robert \surnamestart Glück\surnameend} \&
  \bibinfo{author}{Zhenjiang \surnamestart Hu\surnameend}
  (\bibinfo{year}{2000}): \emph{\bibinfo{title}{A Loop Optimization Technique
  Based on Quasi-Invariance}}, pp. \bibinfo{pages}{80--90}.
\newblock \doi{10.1.1.17.8939}.

\end{thebibliography}
